\documentclass[12pt]{article}
\usepackage{graphicx,amsmath,amssymb}

\newcommand\pubnumber{EFI Preprint 10-25}
\newcommand\pubdate{\today}

\def\address{Enrico Fermi Institute and Department of Physics \\
The University of Chicago, Chicago, Illinois, 60637, USA}
\def\support{\footnote{Work supported by DOE grant DE-FG02-90ER40560.}}

\def\Title#1{\begin{center} {\Large #1 } \end{center}}
\def\Author#1{\begin{center}{ \sc #1} \end{center}}
\def\Address#1{\begin{center}{ \it #1} \end{center}}

\newcommand\pubblock{\rightline{\begin{tabular}{l} \pubnumber\\
         \pubdate  \end{tabular}}}
\newenvironment{Abstract}{\begin{quotation}  }{\end{quotation}}
\newenvironment{Presented}{\begin{quotation} \begin{center} 
             PRESENTED AT\end{center}\bigskip 
      \begin{center}\begin{large}}{\end{large}\end{center} \end{quotation}}
\def\Acknowledgements{  \bigskip \begin{center} \begin{large}
             \bf ACKNOWLEDGEMENTS \end{large}\end{center}}

\begin{document}
\begin{titlepage}
\pubblock

\vfill
\Title{Theory of Inclusive Radiative B Decays}
\vfill
\Author{ Gil Paz\support}
\Address{\address}
\vfill
\begin{Abstract}
This talk discusses recent developments in the theory of inclusive radiative B decays, focusing mainly on non-perturbative aspects of $\bar B\to X_s \gamma$.
\end{Abstract}
\vfill
\begin{Presented}
Proceedings of CKM2010, the 6th International Workshop on the CKM Unitarity Triangle, University of Warwick, UK, 6-10 September 2010
\end{Presented}
\vfill
\end{titlepage}

\section{Introduction}
This talk discusses recent developments in the theory of inclusive radiative B decays, where recent is defined as being after the CKM 2008 workshop. We focus mainly on non-perturbative aspects of $\bar B\to X_s \gamma$, since this is arguably the most important new development in the field.

Inclusive radiative B decays, namely  $\bar B\to X_s \gamma$, $\bar B\to X_d \gamma$, and $\bar B\to X_s l^+l^-$ are important probes of new physics. Since there are no tree-level flavor changing neutral currents in the Standard Model (SM), these processes can only proceed via loop suppressed transitions. These loops are sensitive to the masses and coupling of heavier SM particles, but they can be also sensitive to masses and couplings of particles that  appear in extensions of the SM. By having a good theoretical control of the SM prediction and in particular the non-perturbative effects, one can place important constraints on these models of ``new physics". Currently the experimental value of ${\rm Br}(\bar B\to X_s\gamma,\, E_\gamma> 1.6\, {\rm GeV})=(3.55\pm 0.24\pm 0.09)\cdot 10^{-4}$ \cite{TheHeavyFlavorAveragingGroup:2010qj} is in agreement with the three theoretical predictions, which differ in the way they address the photon cut effects. The first, ${\rm Br}(\bar B\to X_s\gamma,\, E_\gamma> 1.6\, {\rm GeV})=(3.15 \pm 0.23) \cdot 10^{-4}$ \cite{Misiak:2006zs} by Misiak et al., assumes that at the value of  $E_\gamma> 1.6$ GeV one can ignore these cut effects. The second,  ${\rm Br}(\bar B\to X_s\gamma,\, E_\gamma> 1.6\, {\rm GeV})=(2.98\pm 0.26)\cdot 10^{-4}$  \cite{Becher:2006pu} by Becher and Neubert, is obtained by an MSOPE based analysis of the cut related effects. The third,  ${\rm Br}(\bar B\to X_s\gamma,\, E_\gamma> 1.6\, {\rm GeV})=(3.47\pm0.48\pm 0.17)\cdot 10^{-4}$ \cite{Andersen:2006hr} by Andersen and Gardi, is obtained by DGE based analysis of the cut related effects (We have added the second error of 5\% from non-perturbative effects which was not included in \cite{Andersen:2006hr}).

Since the $b$-quark mass, $m_b$, is much larger than $\Lambda_{\rm QCD}$, one can expect that the partonic rate $\Gamma(b\to s \gamma)$ is equal to  $\Gamma(\bar B\to X_s \gamma)$, up to effects suppressed by some power of $\Lambda_{\rm QCD}/m_b$. For a long time it was assumed that just like other inclusive B decays, such as $\bar B\to X_c l \bar\nu$, these effects arise only at the second power of this ratio. We now know that it is not the case. In fact,  non-perturbative effects arise already at order $\Lambda_{\rm QCD}/m_b$. The reason is that unlike  $\bar B\to X_c l \bar\nu$, there is more than one operator in the effective Hamiltonian that can contribute to the decay. As it is well known, apart from  the dipole operator $Q_{7\gamma}=(-e/8\pi^2)m_b \bar s \sigma_{\mu\nu}F^{\mu\nu}(1+\gamma_5)b$ the operators  $Q_{8g}=(-g/8\pi^2)m_b {\bar s} \sigma_{\mu\nu}G^{\mu\nu}(1+\gamma_5){ b}$ and $Q_1^c=(\bar c{ b})_{V-A}({\bar s} c)_{V-A}$ are also important\footnote{Penguin operators contributions  are  suppressed by their small Wilson coefficients.} (see \cite{Benzke:2010js} for notation).  Loosely speaking, we can convert the gluon or the quark-pair into a photon, but at a cost of a perturbative factor of $\alpha_s$ or a non-perturbative factor of  $\Lambda_{\rm QCD}/m_b$. While the former effects are well studied, the latter have received much less attention in the literature.  It is the contribution of operators other than $Q_{7\gamma}$ that give rise to these enhanced non-perturbative effects. Furthermore, unlike non-perturbative contribution familiar from other inclusive B decays, these non-perturbative effects are not given in terms of matrix elements of local operators, but instead in terms of non-local ones. 

Taking these effects into account, the following picture emerges. In the endpoint region of $m_b-2E_\gamma\sim\Lambda_{\rm QCD}$, the photon spectrum can be factorized, symbolically, as
\begin{eqnarray}\label{fact2}
  d\Gamma(\bar B\to X_s\gamma)&=&H\cdot J \otimes S
 +H\cdot J \otimes s\otimes \bar{J}+H\cdot J \otimes s\otimes \bar{J}\otimes \bar{J}\,.
\end{eqnarray} 
 
The first term of (\ref{fact2}) is the ``direct photon" contribution familiar from the factorization formula for $\bar B\to X_u l\,\bar\nu$ in the endpoint region. The direct photon contribution arises from diagrams in which the photon couples directly to the weak vertex. $H$ are hard functions parameterizing physics at the scale $m_b$, $J$ are jet functions describing the physics of the hadronic final state $X$ with invariant mass $M_X\sim\sqrt{m_b\Lambda_{\rm QCD}}$, and $S$ are soft functions incorporating hadronic physics associated with the scale $\Lambda_{\rm QCD}$. 

The second and third terms corresponds to the ``resolved" photon contribution in which the photon couples to light partons instead of coupling directly to the weak vertex. Unlike the direct photon contribution, the resolved photon contribution arises only at order $\Lambda_{\rm QCD}/m_b$ and higher. The second (third) term of (\ref{fact2})  corresponds to the interference of an amplitude in which the photon does not couple directly to the weak vertex  with an amplitude in which the photon couples (does not couple) directly  to the weak vertex.  The new jet functions  $\bar J$ probe the hadronic substructure of the photon at a scale of order $\sqrt{2E_\gamma\Lambda_{\rm QCD}}$. The new soft functions, $s$, describe the soft interactions between the hadronic substructure of the photon and other soft particles, e.g. the heavy quarks. Unlike the soft functions in the direct photon contribution, they contain non-localities in two light-cone directions.  See \cite{Benzke:2010js} for a more detailed discussion.

As one integrates over large enough portions of phase space, the direct photon contribution 
reduces to matrix elements of local operators multiplied by calculable short distance coefficients.  The resolved  photon contribution does not reduce to matrix elements of local operators, but instead to non-local matrix elements convoluted with calculable short distance coefficients. The resolved photon contribution arises only at order $\Lambda_{\rm QCD}/m_b$. Since non-local matrix elements are functions, they are harder to model. This ultimately limits  the precision one can achieve in the theoretical prediction for inclusive radiative B decays and in using them as a tool to constrain new physics. 

The rest of the talk is organized as follows. In section \ref{sec:perturbative} we discuss recent developments that has to do with perturbative aspects of $\bar B\to X_s \gamma$. In section  \ref{sec:np} we discuss recent developments that has to do with non-perturbative aspects of $\bar B\to X_s \gamma$. These have implications also for $\bar B\to X_d \gamma$ which we comment on in section \ref{sec:b2d}. In section \ref{sec:b2sll} we briefly review recent developments in $\bar B\to X_s l^+l^-$. We present our conclusions in section \ref{sec:conclude}.

Two other issues that we do not have time to discuss in detail are resummation and extrapolation.  The first has to do with photon energy cut related effects. We refer the reader to E. Gardi's talk at CKM 2008 \cite{Gardi:2008}.  The second is the fact that theory and experiment are not compared directly. The experimental value of  \cite{TheHeavyFlavorAveragingGroup:2010qj} is \emph{extrapolated} from the measured $E_\gamma\sim 1.9$ GeV to $E_\gamma>1.6$ GeV using \cite{Buchmuller:2005zv}. Considering our new knowledge of non-perturbative effects and the newer Belle measurement with $E_\gamma>1.7$ GeV \cite{:2009qg}, which can potentially allow us to compare these extrapolation factors against data, it is time to revisit the issue.

\section{Recent developments in \boldmath $\bar B\to X_s \gamma$ :\\ Perturbative Aspects}\label{sec:perturbative}
Recent developments in the understanding of of the perturbative aspects of $\bar B\to X_s \gamma$ can be divided into three parts. The continued effort to complete the NNLO prediction for $\Gamma(b\to s \gamma)$,  the calculation of the $(\Lambda_{\rm QCD}/m_b)^2$ corrections to the $Q_{7\gamma}-Q_{7\gamma}$ contribution to the integrated rate at order $\alpha_s$, and the calculation of the subleading jet functions contribution to $Q_{7\gamma}-Q_{7\gamma}$ contribution to the spectrum. In other words, these are all parts of the direct photon contribution to the integrated rate. We discuss each of these in turn. 
\subsection{ \boldmath $\Gamma(b\to s \gamma)$ at NNLO}  
As was discussed in the introduction, $\Gamma(\bar B\to X_s \gamma)=\Gamma(b\to s \gamma)+{\cal O}\left(\Lambda_{\rm QCD}/m_b\right)$. There is a continuing effort to complete the calculation of $\Gamma(b\to s \gamma)$ at NNLO. For details we refer the reader to C. Greub talk at the CKM 2008 \cite{Greub:2008}. Out of the three necessary ingredients: matching at $\mu\sim M_W$, running from $\mu\sim M_W$ to $\mu\sim m_b$, and calculation of the matrix elements at ${\cal O}(\alpha^2_s)$,  the first two are complete and the third is almost done. 

Since CKM 2008 there has been several new calculations that aim to complete the last ingredient. 
The ${\cal O}(\alpha^2_s)$ corrections from $Q_{7\gamma}-Q_{8g}$ were calculated in \cite{Asatrian:2010rq}. 
The authors conclude that this contribution ``..will not alter the central value of  \cite{Misiak:2006zs} by more than 1\%." Details on evaluation of the NNLO QCD corrections in the
heavy charm limit ($m_c \gg m_b/2$) appeared in \cite {Misiak:2010sk}. After the CKM 2010 workshop, the ${\cal O}(\beta_0\alpha^2_s)$ corrections from $Q_{8g}-Q_{8g}$ were calculated in \cite{Ferroglia:2010xe}. The authors conclude that the correction to the branching ratio ``amounts to a relative shift of $+0.12\%$" for the cut of $E_\gamma >1.6$ GeV. The results of  \cite{Ferroglia:2010xe} were confirmed in \cite{Misiak:2010tk}, which also calculated the ${\cal O}(\beta_0\alpha^2_s)$ corrections from $Q_{1}-Q_{8g}$ and $Q_{2}-Q_{8g}$. The authors conclude  that  ``numerical effects of all these	quantities on the branching ratio remain within the $\pm 3\%$ perturbative uncertainty estimated in \cite{Misiak:2006zs}". Finally, as discussed in  \cite {Misiak:2010sk}, the complete ${\cal O}(\alpha^2_s)$ calculation of $Q_1-Q_{7\gamma}$ and $Q_2-Q_{7\gamma}$ is underway, where the goal is  to ``make the perturbative uncertainties... negligible with respect to the non-perturbative... and experimental... ones".

\subsection{\boldmath ${\cal O}(\Lambda^2_{\rm QCD}/m^2_b)$ corrections to $\Gamma_{77}$}
Considering \emph{only} $Q_{7\gamma}-Q_{7\gamma}$, power corrections to the integrated rate are from order 
$\Lambda^2_{\rm QCD}/m^2_b$ suppressed local operators. There are two such possible operators:  
the ``kinetic" operator and the  
``chromomagnetic" operator. The perturbative coefficients of these operators were calculated at tree level long time ago in \cite{Falk:1993dh}. Recently, they were calculated  at ${\cal O}(\alpha_s)$ in \cite{Ewerth:2009yr}, where the authors presented analytical expressions for these coefficients. As for the  numerical impact, they conclude that ``The effect on the $\bar B\to X_s \gamma$ rate is below 1\% for $E_\gamma < 1.8$ GeV."

\subsection{Subleading Jet Function contribution to \boldmath$d\Gamma_{77}$}
Considering \emph{only} the $Q_{7\gamma}-Q_{7\gamma}$ contribution to the spectrum, one has the following  symbolic factorization formula in the endpoint region:
\begin{equation}\label{fact}
d\Gamma_{77}\sim{ H}\cdot { J}\otimes {
S}+\frac{1}{m_b}\sum_{i}\,{ H}\cdot { J}\otimes {
s_i}+\frac{1}{m_b}\sum_{i}\,{H}\cdot
{ j_i}\otimes {S}+\,{\cal
O}\left(\frac{\Lambda^2_{\rm QCD}}{m_b^2}\right)
\end{equation} 
The factorization of the leading power term, ${H}\cdot { J}\otimes {S}$, was proven in  \cite{Korchemsky:1994jb,Bauer:2001yt}.
The factorization of the one of the possible subleading power terms $\sum_{i}\,{ H}\cdot { J}\otimes {s_i}$, namely the subleading shape functions (SSF), was proven in \cite{Lee:2004ja, Bosch:2004cb, Beneke:2004in} 
(see also \cite{Bauer:2001mh}). 

The factorization of the second possible term $\sum_{i}\,{H}\cdot
{ j_i}\otimes {S}$, namely the subleading jet functions, was proven recently  in  \cite{Paz:2009ut}. The subleading jet functions are zero at tree level and they were calculated explicitly at order $\alpha_s$ in \cite{Paz:2009ut}. As a result their contribution is suppressed by $\Lambda_{\rm QCD}/m_b$ and $\alpha_s$.  They are relevant for a high precision extraction of $|V_{ub}|$ using charmless and radiative inclusive B decays.  Since they are part of the direct contribution, they reduce to local operators in the integrated rate.

In summary, the three types of new perturbative calculations  are at NNLO level in $\alpha_s$ and/or $\Lambda_{\rm QCD}/m_b$. They are almost at the theoretical limit, which implies that further improvement seems unlikely. Numerically the new perturbative corrections amount to about  $1\%$  correction for $\Gamma(\bar B\to X_s \gamma)$.
 
\section{\boldmath Recent developments in $\bar B\to X_s \gamma$:\\  Non-Perturbative  Aspects} \label{sec:np}
It was a common misconception in the field to assume that just like other inclusive B decays non-perturbative effects arise at ${\cal O}(\Lambda^2_{\rm QCD}/m^2_b)$ . While this is true for the  $Q_{7\gamma}-Q_{7\gamma}$ contribution, it is no longer true when other operators are included.  Over the years there were hints that ``not all is well" in the study of the $Q_{8g}-Q_{8g}$ \cite{Ali:1995bi, Kapustin:1995fk} contribution and the $Q_1-Q_{7\gamma}$ contribution \cite{Voloshin:1996gw, Ligeti:1997tc, Grant:1997ec, Buchalla:1997ky}. In particular in \cite{Ligeti:1997tc} it was stated, without a proof, that ``There is no OPE that allows one to parametrize non-perturbative effects from the photon coupling to light quarks in terms of B meson matrix elements of local operators." Despite this statement, these effects were thought to be under control or small. There was never a systematic study of these effects. In fact, the uncertainty from $Q_{7\gamma}-Q_{8g}$ \cite{Lee:2006wn} was largely
missed\footnote{Spectator effects from $Q_{7\gamma}-Q_{8g}$ were considered in \cite{Donoghue:1995na} but they were estimated in a {model dependent} way that underestimated their effects.}. The conclusion of \cite{Lee:2006wn} was that  non-perturbative corrections to the integrated rate arise already at ${\cal O}(\Lambda_{\rm QCD}/m_b)$. 

A systematic study of these non-perturbative effects was performed recently in \cite{Benzke:2010js}, which established the factorization formula (\ref{fact2}) for the spectrum in the endpoint region.  The new ingredient is the resolved photon contribution for which the photon does not couple directly to the weak vertex. In the integrated rate the resolved photon contribution does not reduce to matrix elements of local operators.  Schematically,  one finds terms of the form  $\Delta\Gamma\sim \bar J\otimes h$, where $\bar J$ is calculable in perturbation theory and  $h$ is a non-local matrix element. More specifically, for a photon energy cut  $E_\gamma>E_0$ define 
$$
   {\cal F}_E(\Delta) 
   = \frac{\Gamma(E_0) - \Gamma(E_0)|_{\rm OPE}}{\Gamma(E_0)|_{\rm OPE}},
$$
where $\Delta=m_b-2E_0$ and $\Gamma(E_0)|_{\rm OPE}$ is the ``older" calculation.  
Assuming $\Delta\gg\Lambda_{\rm QCD}$
\begin{equation}
\begin{split}
   {\cal F}_E(\Delta) 
   &= \frac{C_1(\mu)}{C_{7\gamma}(\mu)}\,
    \frac{{ \Lambda_{17}}(m_c^2/m_b,\mu)}{m_b}
    + \frac{C_{8g}(\mu)}{C_{7\gamma}(\mu)}\,
    4\pi\alpha_s(\mu)\,\frac{{ \Lambda_{78}^{\rm spec}}(\mu)}{m_b} \\
   &\quad\mbox{}+ \left( \frac{C_{8g}(\mu)}{C_{7\gamma}(\mu)} \right)^2
    \left[ 4\pi\alpha_s(\mu)\,\frac{{ \Lambda_{88}}(\Delta,\mu)}{m_b} 
    - \frac{C_F\alpha_s(\mu)}{9\pi}\,\frac{\Delta}{m_b}\,\ln\frac{\Delta}{m_s} \right]
    + \dots \,,
\end{split}
\end{equation}
where \emph{model independently}, 
\begin{eqnarray}\label{Lambdaij}
   {\Lambda_{17}}\Big(\frac{m_c^2}{m_b},\mu\Big)
   &=& e_c\,\mbox{Re} \int_{-\infty}^\infty {  \frac{d\omega_1}{\omega_1} 
    \left[ 1 - F\!\left( \frac{m_c^2-i\varepsilon}{m_b\,\omega_1} \right)
    + \frac{m_b\,\omega_1}{12m_c^2} \right]} {  h_{17}}(\omega_1,\mu) \,, \nonumber\\
   {\Lambda_{78}^{\rm spec}}(\mu) 
   &=& \mbox{Re}  
    \int_{-\infty}^{\infty} 
{   \frac{d\omega_1}{\omega_1+i\varepsilon} }
    \int_{-\infty}^{\infty} {   \frac{d\omega_2}{\omega_2-i\varepsilon}}\,
    {  h_{78}}^{(5)}(\omega_1,\omega_2,\mu) \,, \nonumber\\
   {\Lambda_{88}}(\Delta,\mu)
   &=& e_s^2 \left[ \int_{-\infty}^{\Lambda_{\rm UV}}\!
   {  \frac{d\omega_1}{\omega_1+i\varepsilon} }
    \int_{-\infty}^{\Lambda_{\rm UV}}\!{ \frac{d\omega_2}{\omega_2-i\varepsilon}}\,
    2 {  h_{88}}^{\rm cut}(\Delta,\omega_1,\omega_2,\mu)\right. \nonumber\\
    &&\left.\hspace{5em}- \frac{C_F}{8\pi^2}\,\Delta \left( \ln\frac{\Lambda_{\rm UV}}{\Delta} - 1 \right)
    \right] .
\end{eqnarray}
$F$ arises from a charm quark loop and its explicit form can be found in \cite{Benzke:2010js}. $\Lambda_{\rm UV}$ is introduced to regularize the convolution integrals, but  ${\Lambda_{88}}(\Delta,\mu)$ is independent of  $\Lambda_{\rm UV}$.
The exact definition of $h_{ij}$ appears in \cite{Benzke:2010js}, but schematically they are (F.T. denotes the Fourier transform), 
\begin{eqnarray}
{ {h}_{88}(\omega_1,\omega_2)}\quad &\mbox{ F.T. of}& \quad\langle\bar B|{\bar b}(0)\cdots {
  s}(un){ \bar s}(r\bar n)\cdots{ b}(0) |\bar
B\rangle\nonumber\\
{ {h}_{17}(\omega_1)}\quad &\mbox{ F.T. of} &\quad\langle\bar B|{\bar b}(0)\cdots { G} (s\bar n) \cdots{ b}(0) |\bar
B\rangle\nonumber\\
{ {h}_{78}(\omega_1,\omega_2)}\quad &\mbox{ F.T. of}& \quad\langle\bar B|{ \bar b}(0)\cdots { b}(0)\sum_q\,e_q\,
    { \bar q}(r\bar n) 
    \cdots { q}(s\bar n) |\bar B\rangle.
\end{eqnarray}
Estimating the non-perturbative error for $\Gamma(\bar B\to X_s \gamma)$ reduces to estimating the size of $\Lambda_{ij}$. Naively, for  $\Lambda_{ij}\sim\Lambda_{\rm QCD}\sim 0.5$ GeV, the effects on the rate can be up to $30\%$. Fortunately, it is possible to constrain $\Lambda_{17}$ and $\Lambda_{78}^{\rm spec}$. We now consider each of these parameters separately. 
\subsection{\boldmath $\Lambda_{17}$}
The soft function ${h}_{17}(\omega_1)$ is an even function of $\omega_1$ 
and its normalization is $2\lambda_2\approx 0.24\, {\rm GeV}^2$. It might be natural to model ${h}_{17}(\omega_1)$ as an exponential or a Gaussian which  leads to an estimate of $-10\,{\rm MeV}<\Lambda_{17}<0\,$MeV. But this is not  a conservative bound, since  ${h}_{17}(\omega_1)$ does not have to be positive. Using the models of  \cite{Benzke:2010js}, one finds  $- 60\,\mbox{MeV} < \Lambda_{17} < 25\,\mbox{MeV}$.  $ \Lambda_{17} $ is smaller than the naive power counting estimate, since part of the contribution  is already included in $\Gamma(E_0)|_{\rm OPE}$ \cite{Voloshin:1996gw, Ligeti:1997tc, Grant:1997ec, Buchalla:1997ky}. It is also suppressed since $F$ is peaked around $1$ GeV, where ${h}_{17}(\omega_1)$  is already becoming small. 
\subsection{\boldmath $ {\Lambda_{78}^{\rm spec}}$}
In order to estimate  $ {\Lambda_{78}^{\rm spec}}$ one can try and use one of the two methods. 
The first is to use the Vacuum Insertion Approximation(VIA). By using Fierz transformation one can write  ${h}_{78}(\omega_1,\omega_2)$ as a product of two B-meson light cone distribution amplitudes (LCDAs). $\Lambda_{78}^{\rm spec}$ depends on the LCDA's inverse moment $\lambda_B$. One has $\Lambda_{78}^{\rm spec}\big|_{\rm VIA}    \in e_{\rm spec}\,[-386\,\mbox{MeV}, -35\,\mbox{MeV} ] ,
$
where $e_{\rm spec}=-1/3$ for $B^0$ or $\bar B^0$, and $e_{\rm spec}=2/3$ for $B^\pm$. 
The second approach is to assume  $SU(3)$ flavor symmetry. In this limit,  $\Lambda_{78}^{\rm spec}$ is determined by the isospin asymmetry $\Delta_{0-}$ in $\bar B\to X_s \gamma$ \cite{Misiak:2009nr}. This asymmetry was measured by BaBar using two different methods. The naive average is $\Delta_{0-}=(-1.3\pm 5.9)\%$. Including $30\%$ $SU(3)$ flavor breaking gives,
$
\Lambda_{78}^{\rm spec} 
   \approx -4.5\,\mbox{GeV}\,(e_{\rm spec}\pm 0.05)\,\Delta_{0-}. 
$
Notice that in the flavor averaged rate one has  $e^{\rm avg.}_{\rm spec}= 1/6$ which is effectively a suppression factor. 

\subsection{\boldmath $\Lambda_{88}$}
In this case one models ${\Lambda_{88}}(\Delta,\mu)$ by 
$ \Lambda_{88}(\Delta,\mu)\approx e_s^2\,\Lambda(\mu)$ where $\Lambda(\mu) $ is positive and taken to be in the range $0<\Lambda(\mu)<1$\,GeV.

\subsection{Total Uncertainty}
Using the above values one has  
${\cal F}_E\big|_{17} \in [-1.7, +4.0]\,\%$ and  ${\cal F}_E\big|_{88} \in [-0.3, +1.9]\,\%$. For ${\cal F}_E\big|_{78} $ one  has $ {\cal F}_E\big|_{78}^{\rm VIA} \in [-2.8, -0.3]\,\% $ or ${\cal F}_E\big|_{78}^{\rm exp} \in [-4.4, +5.6]\,\%$ at $ \mbox{(95\% CL)}$.  ``Scanning" over the ranges one has 
$-4.8\% < {\cal F}_E(\Delta) < +5.6\%$ using VIA for $\Lambda_{78}^{\rm spec}$
or 
$ -6.4\% < {\cal F}_E(\Delta) < +11.5\%$ using $\Lambda_{78}^{\rm spec}$ from $\Delta_{0-}$. The last value reflects the large error on $\Delta_{0-}$. But even if the error on  $\Delta_{0-}$ was zero, one would still have  $-4.0\% < {\cal F}_E(\Delta) < +4.8\%$ in this ``ideal case".

In conclusion, one finds a total uncertainty of about $5\%$. This is also the uncertainty from a previous estimate based on  $Q_{7\gamma}-Q_{8g}$ alone \cite{Lee:2006wn}, with an extra  $50\%$ deviation from VIA. This value is also the uncertainty usually assigned to non-perturbative effects in various SM predictions. While numerically the prediction has not changed, it is now based on a much stronger theoretical basis.  Still to be done are the analysis of the resolved  photon contribution to the CP asymmetry, the spectrum, and the non-perturbative parameters extracted from it \cite{Benzke: in preparation}.

\section{\boldmath Comments on $\bar B\to X_d \gamma$}\label{sec:b2d}
$\bar B\to X_d \gamma$ is analogous to $\bar B\to X_s \gamma$, where usually one only needs to replace 
 $V_{qb}V_{qs}^*$ by $V_{qb}V_{qd}^*$. One important difference is that for this decay  $Q_1^u$ is not CKM suppressed and as a result the $Q_1^u-Q_{7\gamma}$ contribution is not CKM suppressed either . In \cite{Buchalla:1997ky} this contribution was estimated to scale as ${\cal O}(\Lambda_{\rm QCD}/m_b)$, although it was not calculated explicitly. In \cite{Benzke:2010js} the contribution of $Q_1^u-Q_{7\gamma}$ to the CP averaged rate was calculated explicitly. It is given in terms of a convolution of 
$P\,{\dfrac{1}{\omega_1}}$ and ${ h_{17}(\omega_1)}$, i.e. the first term in the first line of (\ref{Lambdaij}). Since the former is odd and the latter is even, the convolution vanishes. This removes the largest source uncertainty and makes $\bar B\to X_d \gamma$ as theoretically clean as $\bar B\to X_s \gamma$ \cite{Hurth:2010tk}.

\section{\boldmath Comments on $\bar B\to X_s\, l^+l^-$}\label{sec:b2sll}
For a more detailed discussion of $\bar B\to X_s\, l^+l^-$ see E. Lunghi talk at CKM 2008 \cite{Lunghi:2008}. Here we limit ourselves to a short review of recent developments.
In the region of low $q^2$, i.e. $q^2\in\,[1...6]\, {\rm GeV}^2$ and when one introduces a cut on the invariant mass of $X_s$, namely  $m_X\leq m_X^{\rm cut}$, $d\Gamma_i$ of $\bar B\to X_s\, l^+l^-$ factorizes similarly to $d\Gamma_{77}$ of $\bar B\to X_s \gamma$ in the endpoint region, see (\ref{fact}). Here $d\Gamma_i$, with  $i=T,A,L$, corresponds to angular decomposition of the triple spectrum (called $H_i$ in \cite{Lee:2008xc}).

In  \cite{Lee:2008xc} the contribution of the ``primary" SSF to $d\Gamma_i$ was calculated. The primary SSF are defined to be the SSF that also contribute to  $\bar B\to X_u l\,\bar\nu$ in the endpoint region. The authors of   \cite{Lee:2008xc} find sizable corrections of the order 5\% to 10\% from the primary SSF corrections. These cause a  shift of $\sim\, -0.05\, {\rm GeV}^2$ to $-0.1\, {\rm GeV}^2$ in the zero of the forward-backward asymmetry. 

In \cite{Bell:2010mg} the two-loop calculation of the hard functions was performed. The authors of  \cite{Bell:2010mg} have observed a significant shift in the zero of the forward-backward  asymmetry going from NLO to NNLO. Including the primary SSF contribution, they find the location of the zero to be at 
$q^2_0=(3.34\,...\,3.40)^{+0.22}_{-0.25}\, {\rm GeV}^2$ for  $m_X^{\rm cut} = (2.0\,...\,1.8)\, {\rm GeV}.
$

Finally, following the completed analysis for $\Gamma(\bar B\to X_s \gamma)$ one should ask what is the effect from ``non-primary" SSF? For example, from soft gluon attachments to the charm-loop diagrams. This point was also  stressed in \cite{Bell:2010mg} and requires further study.

\section{Summary and Outlook} \label{sec:conclude}
Inclusive Radiative B decays is a mature field. New factorization formula for photon spectrum in the endpoint region has been established in \cite{Benzke:2010js}, which includes apart from the direct photon contribution familiar from  $\bar B\to X_u\, l \bar \nu$, also new resolved photon contribution. The resolved photon contribution leads to ${\cal O}(\Lambda_{\rm QCD}/m_b)$ corrections to $\Gamma(\bar B\to X_s\,\gamma)$. From a systematic study of these effects, one finds an irreducible error of $\sim 5\%$ on $\Gamma(\bar B\to X_s \gamma)$. This non-perturbative error is the largest of the errors of the SM prediction, and therefore there is no prospect for reducing the total theoretical error below the $5\%$ level. In the near future we can expect the perturbative error to be reduced below the non-perturbative and the experimental errors, and to have an analysis of the resolved  photon contribution to the CP asymmetry in $\bar B\to X_s \gamma$,  the spectrum, and the non-perturbative parameters extracted from it . For $\bar B\to X_s\, l^+l^-$ the effects of the resolved photon contribution are yet to be calculated.

 Beyond that, further theoretical improvement seems unlikely. From the perturbative side, improvement beyond NNLO seems almost impossible and likely to be unjustified considering the non-perturbative and experimental errors. From the non-perturbative side, we are facing irreducible hadronic uncertainties. This implies that in the very near future we will have the definitive SM theoretical predictions for inclusive radiative B decays in years to come.

\Acknowledgements

I would like to thank the organizers of CKM 2010 for inviting me to give this talk, and Martin Gorbahn for his comments on the manuscript. This work is supported in part by the Department of Energy grant DE-FG02-90ER40560.

\end{document}